\documentclass[12pt]{article}
\usepackage{jheppub}
\usepackage{amsmath}
\usepackage{amssymb,amsfonts}
\usepackage{latexsym}
\usepackage{mathtools}
\usepackage{enumerate}
\usepackage{empheq}
\usepackage{dsfont}
\relax
\renewcommand{\theequation}{\arabic{section}.\arabic{equation}}

\def\be{\begin{equation}}
	\def\ee{\end{equation}}
\def\bea{\begin{eqnarray}}
	\def\eea{\end{eqnarray}}

\renewcommand{\vec}[1]{\boldsymbol{#1}}

\usepackage{amsthm}

\newcommand{\g}{\mathfrak g}

\newcommand{\tr}{\text{tr}}

\title{Code construction and ensemble holography of simply-laced WZW models at level 1}

\author{Nikolaos Angelinos}
\affiliation{Yau Mathematical Sciences Center, Tsinghua University, Beijing 100084, China} 
\emailAdd{angelinosn@mail.tsinghua.edu.cn}
\abstract{We introduce a code construction for Wess-Zumino-Witten (WZW) models associated with simply-laced affine Lie algebras at level 1. The chiral primary fields of these rational CFTs can be parametrized by the elements of the outer automorphism group of the affine algebra, which is isomorphic to the discriminant group $G$ of the root lattice. We show that the classification of even, self-dual codes over the alphabet $G$ is equivalent to the classification of modular-invariant CFTs. 
Each individual CFT is dual to a Chern-Simons theory, after gauging the maximal, non-anomalous subgroup of its 
$1$-form symmetry group specified by the code. 
	We calculate the ensemble average of these CFTs, which is holographically dual to ``CS gravity"-- where the bulk theory is summed over topologies.
	When the alphabet 
	$G$ consists only of elements of square-free order, we explicitly show that this ensemble average reproduces the Poincaré series of the vacuum character, which can be interpreted as the CS path-integral summed only over handlebody topologies. However, when 
	$G$ contains elements of non-square-free order, additional contributions from singular topologies arise. }

\arxivnumber{2503.04055}

\begin{document}

\maketitle

\section{Introduction}

In recent years, it has become increasingly evident that certain low-dimensional models of gravity are dual to an ensemble of boundary theories, rather than a single theory. A prominent early example is JT gravity in two-dimensional spacetime, which is dual to an ensemble of quantum-mechanical models \cite{saad2019jt}. In three dimensions, a key example is ``U(1) gravity"—the perturbative sector of Abelian Chern-Simons (CS) theory summed over handlebodies—which is dual to the ensemble of Narain CFTs \cite{Maloney_2020,Afkhami-Jeddi:2020ezh} (see also \cite{Collier:2021rsn,Cotler:2020hgz,Benjamin:2021wzr,Ashwinkumar:2023jtz,Ashwinkumar:2023ctt,Kames-King:2023fpa,Forste:2024zjt,Perez:2020klz,Datta:2021ftn,Benjamin:2021ygh,Chakraborty:2021gzh,Raeymaekers:2021ypf,Benini:2022hzx,Saidi:2024zdj}).

Recent advances in the code-theoretic formulation of Narain CFTs \cite{Dymarsky:2020qom,Dymarsky:2020bps,Dymarsky:2020pzc,Yahagi:2022idq,Furuta:2022ykh,Henriksson:2022dnu,Angelinos:2022umf,Henriksson:2022dml,Dymarsky:2022kwb,Kawabata:2022jxt,Furuta:2023xwl,Alam:2023qac,Kawabata:2023iss,Ando:2024gcf,Barbar:2023ncl,Aharony:2023zit,Dymarsky:2024frx,Kawabata:2025hfd} have enabled a systematic study of discrete subsets of the Narain moduli space, which admit a holographic description in terms of Abelian CS theory. According to the picture developed in \cite{Barbar:2023ncl}, the topological boundary conditions for 3D Abelian CS theory (which correspond to modular-invariant boundary CFTs), are parametrized by even, self-dual codes. Averaging over the boundary theories is holographically dual to coupling the CS theory to topological gravity, where the bulk path-integral is summed over all topologies, including singular ones \cite{Dymarsky:2024frx}.
In the limit of large genus, the path integral of CS summed over handlebodies matches the ensemble average of all boundary theories with uniform weights. The torus partition function can then be obtained using the method of genus reduction. In many cases, this reduction yields a bulk sum over handlebody topologies only. In general, however, this procedure introduces contributions from singular topologies. Similarly, for other rational CFTs (RCFTs) \cite{Castro:2011zq,Jian:2019ubz,Romaidis:2023zpx,Meruliya:2021utr,Meruliya:2021lul}, summing over genus-one handlebodies alone may yield non-physical modular invariants with negative densities of states, necessitating the inclusion of contributions from singular topologies to obtain a physically consistent boundary dual.

Motivated by these developments, we investigate the conditions under which the ensemble average of torus RCFTs can be interpreted as a bulk sum over handlebodies. We focus on ensembles of flavored partition functions of simply-laced Wess-Zumino-Witten (WZW) models at level 1. A key component of our approach is a code-theoretic formulation of these RCFTs.  The primary fields are naturally parametrized by elements of the discriminant group 
$G$ of the corresponding root lattice. The conditions of 
$S$- and 
$T$-invariance for the CFT partition function translate into self-duality and evenness conditions for codes over the alphabet 
$G$. Consequently, modular invariant CFTs are in one-to-one correspondence with enumerator polynomials of even, self-dual codes over 
$G$. The CFT partition functions can be obtained from these polynomials by substituting the arguments for the characters of the relevant representations. While the classification of modular invariants for these models is well known \cite{Degiovanni:1989ne,Itzykson1988LevelOK,Gannon:1992nq}, our approach recasts them in the language of codes, providing an explicit construction in the framework of \cite{Barbar:2023ncl}.

The formulation in terms of additive codes is particularly straightforward due to the underlying free-boson description of these models, which correspond to Narain CFTs at special points of enhanced symmetry. Thus, our analysis includes the construction and ensemble average of a discrete subset of Narain CFTs with fixed enhanced symmetry. The code-based construction of these CFTs was first described in \cite{Angelinos_thesis}. A similar approach, framed in terms of codes over cyclotomic integers, was independently developed in \cite{Mizoguchi:2024ahp}. In this paper we build on \cite{Angelinos_thesis} by describing in detail how this code formulation works and explaining its holographic origin.

The paper is organized as follows. In Section \ref{sec:preliminaries}, we introduce the code formalism used throughout our analysis and discuss its bulk interpretation in terms of Abelian CS theory. In Sections \ref{sec:an}, \ref{sec:dn}, and \ref{sec:en}, we apply this framework to different Lie algebras. Each section follows a similar structure: we classify and enumerate all modular-invariant CFT partition functions using the code description, compute their ensemble average, and discuss its bulk interpretation. We conclude in Section \ref{sec:conclusions}.

\section{Code description of simply-laced affine algebras at level $1$}\label{sec:preliminaries}

Consider a simply-laced Lie algebra $\g$ of rank $r$ and its root lattice $\Lambda$. By abuse of notation, we also use $\Lambda$ to refer to the generator matrix of the root lattice, whose columns are the simple roots. For a simply-laced algebra, the dual root lattice, given by $\Lambda^\perp=(\Lambda^{-1})^T$, is equal to the weight lattice, whose columns contain the fundamental weights $\hat\omega_1,\hat\omega_2,\dots,\hat\omega_{r}$. 

In the affine extension $\g_k$ there is an additional fundamental weight, denoted by $\hat\omega_0$, corresponding to the basic (vacuum) representation. An arbitrary weight $\lambda$ can be expressed as an integer linear combination of fundamental weights $\lambda=\sum_{i=0}^{r} \ell_i \hat\omega_i$, where $\ell_i$ are the Dynkin labels. The integrable highest-weight representations of $\g_k$ correspond to dominant weights, which are weights with nonnegative integer Dynkin labels. 
At a fixed level $k$, the only dominant weights allowed are those satisfying $\sum_{i=1}^r a_i \ell_i\leq k$, where $a_i$ is the comark (equal to the mark, for simply-laced algebras) associated with the $i$-th simple root.  Consequently, at level $k=1$, the only allowed dominant weights are the fundamental weights whose corresponding simple root has comark equal to $1$.

Next, define the discriminant group of the root lattice $G=\Lambda^\perp/\Lambda$, which is a finite Abelian group.  This group is isomorphic to the outer automorphism group $\mathcal O_\g$ of $\g_k$, as well as to the center $\mathcal Z(\mathcal G)$ of the Lie group $\mathcal G$ generated by $\g$.  $\mathcal O_\g$ maps a fundamental weight to another with the same comark. Its action on the Dynkin labels is given in table \ref{table1}. The fundamental weights of unit comark have a single orbit under $\mathcal O_\g$. Since $\hat\omega_0$ always has unit comark, its orbit consists of all the dominant weights at $k=1$. We can, therefore, label the dominant weights (and thus the integrable highest-weight representations) at $k=1$ with elements of $G$.

Now, let  $\phi$ be a surjective homomorphism $\phi:\Lambda^\perp\to G$ with $\ker(\phi)=\Lambda$. For each $g\in G$, the equivalence class $\phi^{-1}(g)$ contains exactly one fundamental weight\footnote{The rest of the elements of $\phi^{-1}(g)$ correspond to descendants of $\omega_g$ under $\g_k$ (and are Virasoro highest-weight representations).} of unit comark, which we denote by $\omega_g\equiv A_g(\hat\omega_0)$, where $A_g\in\mathcal O_\g$ is the outer automorphism corresponding to $g\in G$. This establishes a natural identification between fundamental weights $\omega_g$ of unit comark and elements of $G$. Under these definitions, an outer automorphism $A_{g'}$ acts on $\omega_g$ as $A_{g'}(\omega_g)=\omega_{g+g'}$.

 The group $G$ naturally inherits a bilinear form from the Euclidean inner product on $\Lambda^\perp\subseteq\mathbb R$. Specifically, for $g_1,g_2\in G$ we have
\be \langle g_1|g_2\rangle\equiv \lambda_1\cdot\lambda_2\mod \mathbb Z,\label{Gform}\ee
where $\lambda_1,\lambda_2$ are any elements of $\Lambda^\perp$ such that $g_1=\phi(\lambda_1)$ and $g_2=\phi(\lambda_2)$ and $\cdot$ is the Euclidean dot product. 
It is also useful to define the weight of $g\in G$, as follows
\be wt(g)\equiv\min |\phi^{-1}(g)|^2=\min_{k\in\mathbb Z^{r} }|\Lambda k+\lambda_g |^2={\omega_g\cdot(\omega_g+2\rho)\over 1+h^\perp},\label{Gweight}\ee
where $\lambda_g$ is an element of $\Lambda^\perp$ such that $\phi(\lambda_g)=g$, $h^\perp$ is the dual Coxeter number and $\rho$ is the Weyl vector (obtained by adding all columns of $\Lambda^\perp$).

A \textit{code} is a subgroup of $G^n$. The function $\phi$ can be naturally extended to $\phi:\oplus_n\Lambda^\perp\to G^n$, but there are multiple ways to extend the bilinear form (\ref{Gform}).
  In this paper, we focus on  codes\footnote{In the notation above we use $G^n\times\bar G^n$, rather than $G^{2n}$, to emphasize the negative sign in the bilinear form (\ref{form}).}   $C\subseteq G^n\times\bar G^n$ of Lorentzian signature $(n,n)$, equipped with the bilinear form 
 \be \langle (\vec a,\vec b)|(\vec a',\vec b')\rangle=\sum_{i=1}^n\langle a_i|a_i'\rangle-\sum_{i=1}^{n}\langle b_i|b_i'\rangle,~~(\vec a,\vec b),(\vec a',\vec b')\in G^n\times\bar G^n.\label{form}\ee
 With respect to this bilinear form, the dual code of $C$ is
 \be C^\perp=\{(\vec a,\vec b)\in G^n\times\bar G^n:\langle (\vec a,\vec b)|(\vec a',\vec b')\rangle=0\text{ for all }(\vec a',\vec b')\in C\}.\ee
  A code $C$ is \textit{self-dual} if $C=C^\perp$ and it is \textit{even} when all $c=(\vec a,\vec b)\in C$ satisfy the condition
 \be \text{evenness condition: } wt(\vec a)-wt(\vec b)\equiv\sum_{i=1}^n wt(a_i)-\sum_{i=1}^nwt(b_i)=0\mod 2\mathbb Z.\label{evenness}\ee
The code \textit{enumerator polynomial} is defined as
\be W_{\mathcal C}\equiv \sum_{(\vec a,\vec b)\in\mathcal C}x_{\vec a}\bar x_{\vec b}=\sum_{(\vec a,\vec b)\in G^n\times\bar G^n}M_{\vec a,\vec b}x_{\vec a}\bar x_{\vec b},\label{code_enum}\ee
where we defined
\be x_{\vec a}\bar x_{\vec b}=\prod_{g\in G} x_g^{e_g(\vec a)}\bar x_g^{e_g(\vec b)},\label{eg}\ee
and $e_g(\vec a)$ counts the entries of $\vec a$ that are equal to $g$. The $M_{\vec a,\vec b}$ are non-negative integers counting the multiplicities of the codewords, and $M_{\vec 0,\vec 0}=1$. The MacWilliams transformation relates the polynomial of a code with the polynomial of its dual code as follows
\be W_{\mathcal C^\perp}={1\over |G|^n}\sum_{(\vec a,\vec b)\in G^n\times \bar G^n}\sum_{(\vec a',\vec b')\in G^n\times \bar G^n}M_{\vec a,\vec b}e^{-2\pi i (\langle \vec a|\vec a'\rangle-\langle \vec b|\vec b'\rangle)}x_{\vec a'}\bar x_{\vec b'} \label{MacWilliams}.\ee
In particular, the enumerator polynomial of a self-dual code is invariant under the MacWilliams transformation.

\begin{table}\caption{}\centering
	\begin{tabular}{ |c|c|c|c| }
		\hline
		\text{$\ell$} &$G$ & Bilinear form on $G$ & Action of generators on Dynkin labels  \\
		\hline
		$A_{n-1}$ & $\mathbb Z_n$ &$\langle g|g'\rangle$=${n-1\over n}gg'$ &  $[\ell_0,\ell_1,\cdots,\ell_{n-1}]\to[\ell_1,\ell_2,\cdots,\ell_{n-1},\ell_0] $ \\ 
		$D_{n=2l}$ & $\mathbb Z_2\otimes \mathbb Z_2$ &$\langle (g_1,g_2)|(g_1',g_2')\rangle$=${n\over 4}(g_1g_1'+g_2g_2')$ & $[\ell_0,\ell_1,\cdots,\ell_{n}]\to[\ell_1,\ell_0,\ell_2\cdots,\ell_{n},\ell_{n-1}] $ \\ 
		& &  ~~~\quad\quad\quad\quad\quad$+{n-2\over 4}(g_1g_2'+g_1'g_2)$ & $[\ell_0,\ell_1,\cdots,\ell_{n}]\to[\ell_{n},\ell_{n-1},\ell_{n-2},\cdots,\ell_{1},\ell_0] $  \\
		$D_{n=2l+1}$ & $\mathbb Z_4$ &$\langle g|g'\rangle$=${n\over 4}gg'$ & $[\ell_0,\ell_1,\cdots,\ell_{n}]\to[\ell_{n-1},\ell_{n},\ell_{n-2},\dots,\ell_{1},\ell_0] $ \\ 
		$E_6$ &$\mathbb Z_3$& $\langle g|g'\rangle$=${4\over 3}gg'$ & $[\ell_0,\ell_1,\cdots,\ell_{6}]\to[\ell_{1},\ell_5,\ell_{4},\ell_3,\ell_6,\ell_{0},\ell_2] $ \\
		$E_7$ &$\mathbb Z_2$& $\langle g|g'\rangle$=${3\over 2}gg'$ &  $[\ell_0,\ell_1,\cdots,\ell_{7}]\to[\ell_{6},\ell_5,\ell_{4},\ell_3,\ell_2,\ell_{1},\ell_0,\ell_7] $ \\
		$E_8$ &\text{trivial}& $-$ & $-$  \\
		\hline
	\end{tabular}\label{table1}
\end{table}

 \subsection{Affine characters and their modular transformations}

 The central charge of the $\g_1$ WZW model is equal to the rank of $\g$, as determined by the Sugawara construction
\be c={\dim \g\over 1+h^\perp}=r,\ee
where we used that $\dim\g=(1+h^\perp)r$ for simply-laced algebras.
At $k=1$, there are $|G|$ primary fields, corresponding to the dominant weights $\omega_g$.  The conformal dimension $h_g$ of $\omega_g$ is given by
 \be h_g={\omega_g\cdot(\omega_g+2\rho)\over 2(1+h^\perp)}={wt(g)\over 2},~g\in G.\ee
 We define the flavored characters of $\g_1$ as follows \cite{DiFrancesco:1997nk}
  \be \chi_g(\tau;\xi,t)\equiv \tr_{\omega_g}\big[e^{-2\pi i\hat k t}e^{2\pi i\tau (L_0-c/24)}e^{-2\pi i\xi\cdot H}\big],\label{chartr}\ee
  where the trace is over the module of highest weight $\omega_g$, $H^i$ are the Cartan generators in the Cartan-Weyl basis and $\hat k$ is the central element, which has eigenvalue $1$ in our case.
For a simply-laced algebra at level $1$, they can be written as
 \be \chi_g(\tau;\xi,t)= e^{-2\pi i t}{1\over (\eta(\tau))^{r}}\sum_{n\in \mathbb Z^{r}}e^{\pi i\tau (\Lambda n+\lambda_g)^T(\Lambda n+\lambda_g)}e^{-2\pi i \xi\cdot(\Lambda n+\lambda_g)}.\label{char}\ee
Underlying this simplified form of the characters is the fact that simply-laced $\g_1$ WZW models have an equivalent description in terms of a Narain theory of $r$ compact free bosons $\varphi^j$. The generators of the Cartan algebra are identified with $H^j=i\partial\varphi^j$, while the ladder operators, parametrized by the roots $\alpha$, are vertex operators $E^\alpha\sim e^{i\alpha\cdot\varphi}$. The $u(1)^r$ characters can be organized into the $\g_1$ characters, resulting in \eqref{char}.

 Under the modular group, the characters transform as follows\footnote{We chose to exclude the phase $e^{-r{\pi i\over 12}}$ from the definition of the $T$ matrix, since it cancels out upon combining holomorphic and anti-holomorphic parts.}
 \be \chi_g(\tau+1;\xi,t)=e^{-r{\pi i\over 12}}\sum_{g'\in G}T_{gg'}\chi_{g'}(\tau;\xi,t),~~T_{gg'}=\delta_{gg'}e^{\pi i \;wt(g)},\label{T}\ee
 \be \chi_g(-1/\tau;\xi/\tau,t+{\xi^2\over 2\tau})=\sum_{g'\in G}S_{gg'}\chi_{g'}(\tau;\xi,t),~~S_{gg'}={1\over \sqrt{|G|}}e^{-2\pi i\langle g|g'\rangle},\label{S}\ee
 where the bilinear form $\langle g|g'\rangle$ for each algebra is written explicitly in table \ref{table1}.

 A straightforward application of the Verlinde formula on the $S$ matrix leads to fusion numbers $N^c_{ab}=\delta_{c,a+b}$, i.e. the fusion rules are
 \be [\omega_{g}]\times [\omega_{g'}]=[\omega_{g+g'}],~~~g,g'\in G.\ee
 
 \subsection{Partition functions and code polynomials}
The torus partition function of the CFT is a modular invariant combination of the characters
 \be Z=\sum_{(\vec a,\vec b)\in G^n\times\bar G^n} M_{\vec a,\vec b}\chi_{\vec a}\bar \chi_{\vec b},\label{Zi}\ee
 where $M_{\vec a,\vec b}$ are non-negative integers, with $M_{\vec 0,\vec 0}=1$. The classification of all modular invariants at level $1$ is well-known \cite{Degiovanni:1989ne,Itzykson1988LevelOK,Gannon:1992nq}. In this section we show that for simply-laced algebras, the classification of these invariants is equivalent to the classification of even, self-dual codes in $G^n\times\bar G^n$.

  Under the $S,T$ generators of the modular group, using (\ref{T}) and (\ref{S}) we have
 \be  T:Z\to Z'= \sum_{\vec a,\vec b} e^{\pi i (wt(\vec a)-wt(\vec b))} \chi_{\vec a}\bar\chi_{\vec b} M_{\vec a,\vec b},\ee
 \be S:Z\to Z'={1\over |G|^n}\sum_{\vec a,\vec a',\vec b,\vec b'} e^{-2\pi i (\langle \vec a'|\vec a\rangle-\langle \vec b'|\vec b\rangle)}\chi_{\vec a'}\bar\chi_{\vec b'}M_{\vec a',\vec b'}.\label{SZi}\ee
 Invariance under the $T$ transformation is equivalent to requiring that all tuples $(\vec a,\vec b)$ in (\ref{Zi}) with $M_{\vec a,\vec b}\neq 0$ obey the evenness condition \eqref{evenness}. From (\ref{SZi}) we see that the $S$ transformation acts on the characters in the same manner as the MacWilliams transformation (\ref{MacWilliams}) acts on code enumerator variables, hence $S$-invariance is equivalent to requiring that all tuples $(\vec a,\vec b)$ appearing in (\ref{Zi}) belong to a self-dual code.
 Therefore, classifying all modular invariant combinations is equivalent to classifying all even, self-dual codes $C\subseteq G^n\times \bar G^n$. 
 The partition function is obtained from the code enumerator polynomial (\ref{code_enum}) by the substitution
 \be x_g\to \chi_g(\tau;\xi,t),~\bar x_{g}\to \bar \chi_g(\bar\tau;\bar\xi,\bar t).\label{polytochar}\ee
 From now on, we will use code variables $x_g$ and characters $\chi_g(\tau;\xi,t)$ interchangeably (as well as enumerator polynomials $W$ and CFT partition functions $Z$), keeping in mind the correspondence (\ref{polytochar}).
 
 \subsection{Narain description and the bulk picture}
 
 The simply-laced $\g_1$ WZW models are equivalent to a theory of $r$ free bosons, compactified on a specific lattice. Due to this equivalence, we can describe the bulk dual theory in terms of Abelian Chern-Simons.
 The Narain lattice  $\mathcal L_C$ can be obtained by applying the generalized construction A \cite{Angelinos:2022umf} to the even, self-dual code $C\subseteq G^n\times\bar G^n$ as follows
\be \mathcal L_C=\{l\in \underbrace{\Lambda^\perp\oplus\cdots\oplus\Lambda^\perp}_{2n\text{ terms}} :\phi(l)\in C\}.\label{consA}\ee
The bulk description of these Narain CFTs is given in terms of $U(1)^{nr}\times U(1)^{nr}$ Chern-Simons theory on a 3d handlebody $M$
\be \mathcal S={iK_{ij}\over 4\pi}\int_{M} (A^i\wedge dA^j-B^i\wedge dB^j),\label{CS}\ee
where $K=\oplus_{i=1}^n\Lambda^T\Lambda$ is the Cartan matrix of the semi-simple Lie algebra $\oplus_n\g$. The distinct, gauge-invariant Wilson lines $\mathcal W_{\vec a,\vec b}(\gamma)$ are parametrized by non-contractible loops $\gamma$ and $(\vec a,\vec b)\in G^n\times\bar G^n$ \cite{Belov:2005ze}, which is the 1-form symmetry group \cite{Gaiotto:2014kfa} of this theory. The group $G^n\times\bar G^n$ describes the fusion of anyons, corresponding to these Wilson lines.
 
 The path integral of the CS theory on a handlebody $M$ defines a state on its boundary. For a torus, the Hilbert space $\mathcal H$ of boundary states has dimension $|G^n\times \bar G^n|$. A basis can be constructed by inserting Wilson lines $\mathcal W_{\vec a,\vec b}(\gamma)$ in the path integral with all possible charges $(\vec a,\vec b)\in G^n\times \bar G^n$ winding around the non-contractible cycle $\gamma$ of the handlebody $M$ bounded by the torus. A non-contractible line $\mathcal W_{\vec a,\vec b}(\gamma)$ gives rise to the conformal block corresponding to the code monomial $x_{\vec a}\bar x_{\vec b}$. 
 
 Clearly, the blocks obtained this way are not modular invariants. To obtain a full-fledged CFT (and thus a modular invariant CFT partition function), one needs to gauge a maximal, non-anomalous subgroup of the 1-form symmetry group \cite{Benini:2022hzx}. A non-anomalous subgroup $C\subseteq G^n\times\bar G^n$ is one containing lines, parametrized by $(\vec a,\vec b)\in C$, such that their spin and pairwise braidings are trivial. These conditions are equivalent to $C$ being an even, self-dual code \cite{Barbar:2023ncl}. The spin statistics of the anyon parametrized by $(\vec a,\vec b)\in G^n\times\bar G^n$ are described by the phase $\theta(\vec a,\vec b)$ it acquires after a Dehn twist of the torus
 \be \theta(\vec a,\vec b)=\prod_{i=1}^nT_{a_ia_i}T_{b_ib_i}^\star=e^{\pi i(wt(\vec a)-wt(\vec b))},\ee
 where $T$ is defined in (\ref{T}), while the braiding of two lines is given by \cite{Kaidi:2021gbs}
 \be B((\vec a,\vec b),(\vec a',\vec b'))={\theta(\vec a+\vec a',\vec b+\vec b')\over \theta(\vec a,\vec b)\theta(\vec a',\vec b')}=e^{2\pi i (\langle \vec a|\vec a'\rangle-\langle \vec b'|\vec b\rangle)}.\ee
Clearly,  $C$ must be an even, self-orthogonal code. Additionally, the requirement that $C$ is maximal, i.e. there exists no anyon outside of $C$ with trivial braiding with all anyons in $C$, means that $C$ is self-dual. 
 Gauging the subgroup $C$ is equivalent to summing over all insertions of Wilson lines in $C$ \cite{Gaiotto:2014kfa}. This results in a trivial CS theory with a unique state, that is modular invariant. This modular-invariant state  corresponds to the code enumerator polynomial
 \be W_C=\sum_{(\vec a,\vec b)\in C} x_{\vec a} x_{\vec b}.\ee
 Equivalently, this modular-invariant state can be prepared from the path-integral of (\ref{CS})  on the manifold $M=T^2\times [0,1]$ (where two tori are connected by an interval), by imposing topological boundary conditions described by the code $C$ at $T^2\times\{1\}$ \cite{Barbar:2023ncl}. This configuration can be ``unfolded" into a $U(1)^{nr}$ CS theory (only the $A$ fields in (\ref{CS})) on $T^2\times[0,2]$, with a surface operator \cite{KS1,KS2,Roumpedakis:2022aik} described by the code $C$  inserted at $T^2\times \{1\}$.

As proposed in \cite{Dymarsky:2024frx}, we now consider the average over the ensemble of all maximal gaugings of 1-form symmetries of (\ref{CS}), i.e. the average over all even, self-dual codes. This process is holographically dual to coupling the bulk CS theory to topological gravity.
 We define the CFT partition function averaged over this ensemble with equal weights
\be \langle W\rangle={1\over \mathcal N}\sum_i W_{C_i},\ee
where $i$ runs over all even, self-dual codes and $\mathcal N$ is a normalization constant.
 This expression is manifestly modular invariant and can be expressed as a
 Poincar\'e series
\be \langle W\rangle\propto 
 \sum_{\gamma\in SL(2,\mathbb Z)/\Gamma}\gamma(X_\text{seed}),\label{seed}\ee
 with some appropriate seed, where $\Gamma$ is a subgroup of $SL(2,\mathbb Z)$ acting trivially on $X_{\text{seed}}$.
 
 In some cases, the ensemble average is equal to the Poincar\'e series of the vacuum character $X_\text{seed}=x_{\vec 0}\bar x_{\vec 0}$, suggesting a simple, semi-classical ``TQFT gravity"  interpretation, where the bulk TQFT is summed only over handlebody topologies.
It is therefore useful to define $\hat W$ as the Poincar\'e series of the vacuum character
 \be \hat W\equiv \sum_{\gamma\in SL(2,\mathbb Z)/\Gamma}\gamma(x_{\vec 0}\bar x_{\vec 0}),\label{pseries}\ee
 where $\Gamma$ is a subgroup of $SL(2,\mathbb Z)$ that acts trivially on the vacuum character. For a rational CFT, $\Gamma$ is a finite-index subgroup of the modular group, resulting in finitely many classes of handlebodies that contribute to this sum. One of the main goals of the subsequent sections will be to examine whether $\langle W\rangle$ is proportional to $\hat W$.

 \section{The $su(N)^n_1$ WZW models}\label{sec:an}

 We begin with the $su(N)_1$ algebra, which exhibits the most interesting structure.   The Poincar\'e series of this theory at $n=1$ has been discussed earlier by \cite{Meruliya:2021utr}. Our analysis, framed in the language of codes, generalizes their results for $n>1$ and for flavored characters. Including flavor has the benefit of making the characters (and the partition functions) linearly independent and the $S$ matrix well-defined through equation (\ref{S}).
 
 The $su(N)$ algebra has rank $r=N-1$ and all comarks equal to $1$, meaning that there are $N$ dominant weights at level $1$.
 A generator matrix for its root lattice $\Lambda=A_{N-1}$ is given by
 \be \Lambda_{ij}=\begin{cases}
 	\sqrt{i+1\over i} & i=j\\
 	-\sqrt{i\over i+1} & j=i+1\\
 	0 & \text{otherwise}.
 \end{cases},~~i,j=1,2,\dots,N-1\ee
 The discriminant group is
  \be G=\Lambda^\perp/\Lambda\cong \mathbb Z_N,\ee
  whose elements we use to label the dominant weights.
  The following weight $\lambda\in\Lambda^\perp$ always has order $N$ in the quotient $\Lambda^\perp/\Lambda$
 \be \lambda=(0,0,\dots,0,\sqrt{N-1\over N})^T,\ee
thus we choose the map $\phi:\Lambda^\perp\to \mathbb Z_N$ such that $\phi(\lambda)=1$.
 In other words, the inverse map $\phi^{-1}$ acting on an element $a$ of $\mathbb Z_N$ results to the following set of vectors in $\Lambda^\perp\subseteq \mathbb R^{N-1}$:
 \be a\mapsto \left\{\Lambda m+a\lambda:~m\in\mathbb Z^{N-1}\right\}.\ee
With this choice of $\phi$, the bilinear form \eqref{Gform} on $G$ reads
 \be \langle a_1|a_2\rangle=a_1a_2|\lambda|^2={N-1\over N}a_1a_2\mod\mathbb Z,\label{inner_productG}\ee
 while the group weight (\ref{Gweight}) is
 \be wt(a)\equiv \min_{k\in\mathbb Z^{N-1}} ||\Lambda k+a\lambda ||={a(N-a)\over N},~a=0,1,\dots,N-1.\ee
 This results in the conformal dimensions
 \be h_a={wt(a)\over 2}={a(N-a)\over 2N},~a=0,\dots,N-1.\ee 
From (\ref{evenness}) we also obtain the evenness condition  on the codes $C\subseteq G\times\bar G$
 \be \text{evenness condition: }{N-1\over N}(a^2-b^2)=0\mod 2\mathbb Z.\label{evenan}\ee
 The modular $T,S$ matrices are explicitly given by:
 \be T_{aa'}=\delta_{aa'}e^{\pi i (N-1){a^2\over N}},~~S_{aa'}={1\over \sqrt{N}}e^{-2\pi i{aa'\over N}}.\label{ANTmatrix}\ee
 
 \subsection{Classification of codes of length $n=1$}

In this subsection we enumerate all the even, self-dual codes of length $n=1$. We begin with the case $N={p^{m}}$, where $p$ is prime and $m$ is a positive integer. Each self-dual code with alphabet $\mathbb Z_{p^m}$ is isomorphic, as an additive group, to $\mathbb Z_{p^{m-k}}\times\mathbb Z_{p^{k}}$ for $k=0,1,\dots,m$, resulting in $\sigma_0(p^m)=m+1$ self-dual codes, where $\sigma_0(a)$ denotes the number of divisors of $a$. 
 Their generators are\footnote{ For even $m$ and odd prime $p$, the code $\mathcal C_{m\over 2}$ is a direct sum of two codes of length $m/2$, with factorizable enumerator polynomial $W=\sum_{a=0}^{p^{m/2}-1} x_{a p^{m/2}}\sum_{a=0}^{p^{m/2}-1} \bar x_{a p^{m/2}}$. This happens because $p^m-1$ is a multiple of $8$, a dimension where even, self-dual Euclidean lattices exist.}
 \be \mathcal C_0=(1,1),~\mathcal C_m=(1,-1),~\mathcal C_k=\begin{cases}\begin{pmatrix}
 		p^{k} & p^{k}\\
 		0 & p^{m-k}
 	\end{pmatrix} & 1\leq k\leq{m\over 2},\\
 	\begin{pmatrix}
 		p^{m-k} & -p^{m-k}\\
 		0 & p^{k}
 	\end{pmatrix}& {m\over 2}<k\leq m-1\end{cases}.\label{asq}\ee
  For odd $p$, it is clear from (\ref{evenan}) that each self-dual code is automatically even.
 For $p=2$ and even $m$, the code $\mathcal C_{m/2}$ is not even. If $p=2$ and $m$ is odd, the codes $\mathcal C_{(m-1)/2}$ and $\mathcal C_{(m+1)/2}$ are identical. In either case, this decreases the number of even, self-dual codes for $p=2$ to $\sigma_0(2^{m}/2)=m$.
 
 Now consider general $N$ with prime factorization $N=\Pi_{i=1}^{q} p_i^{m_i}$. By the Chinese Remainder Theorem (CRT), there exists an isomorphism 
 \be \pi: \mathbb Z_N\to \bigotimes_{i=1}^q \mathbb Z_{p_i^{m_i}}.\label{CRT}\ee
 Let $\mathcal D_i$ denote an even, self-dual code with alphabet $\mathbb Z_{p_i^{m_i}}$. Given the collection $\{\mathcal D_i,i=1,\dots,q\}$, we can construct an even, self-dual code $\mathcal C$ with alphabet $\mathbb Z_N$ by combining the product code $\otimes_i \mathcal D_i$ under the map $\pi^{-1}$  \cite{crtcodes}. Conversely, any even, self-dual code $\mathcal C$ over $\mathbb Z_N$ can be decomposed into a family of even, self-dual codes $\{\mathcal D_i,i=1,\dots,q\}$, each over $\mathbb Z_{p_i^{m_i}}$. This leads to an one-to-one correspondence between even, self-dual codes over $\mathbb Z_N$ and collections of even, self-dual codes $\{\mathcal D_i,i=1,\dots,q\}$, each over a factor $\mathbb Z_{p_i^{m_i}}$. Counting the latter is straightforward, leading to
 \be \text{number of even, self-dual codes}~~\kappa_N\equiv\begin{cases}
 	\sigma_0(N)& N\text{ odd}\\
 	\sigma_0(N/2)& N\text{ even}.
 \end{cases}\label{numberofcodes}\ee
Even, self-dual codes can be obtained by ``orbifolding" the diagonal code $\mathcal C_0$ by subgroups of $\mathbb Z_N$ (see appendix \ref{appOrbifold}). There are $\sigma_0(N)$ subgroups of $\mathbb Z_N$, however for even $N$ the subgroups of odd index must be excluded, since they do not result in distinct even codes. This leads to a counting in agreement with \eqref{numberofcodes}.

The enumerator polynomial $W(x,\bar x)$ of $\mathcal C$ can be obtained from the product of the enumerator polynomials $W^{(i)}(x^{(i)},\bar x^{(i)})$ of $\mathcal D_i$. First define the action of (\ref{CRT}) on the enumerator polynomial variables $\pi^{-1}(x^{(1)}_{g_1}x^{(2)}_{g_2}\cdots x^{(q)}_{g_q})=x_{\pi^{-1}(g_1,g_2,\dots,g_q)}$. Then, the polynomial of $\mathcal C$ is $W(x,\bar x)=\pi^{-1}(\prod_i W^{(i)}(x^{(i)},\bar x^{(i)}))$. An explicit formula for the enumerator polynomials of even, self-dual codes is given by \cite{Degiovanni:1989ne}
\be W_\delta(x,\bar x)=\sum_{a\in \mathbb Z_{N/n_\delta}}\sum_{b\in \mathbb Z_{n_\delta}}x_{a n_\delta}\bar x_{a s_\delta+b N/n_\delta} ,\label{Wd}\ee
where $\delta$ is a divisor of $N$, $n_\delta=\gcd(\delta,N/\delta)$ and $s_\delta=q_1{N\over \delta }+q_2{\delta} \mod{{N\over n_\delta}}$, where $q_1,q_2$ are any two integers satisfying $q_1{N\over \delta }-q_2{\delta}=n_\delta$. We emphasize again that for even $N$, the choices of $\delta$ containing a single factor of $2$ must be excluded.

 The specialized characters $\chi_g(\tau;0,0)$ are symmetric under charge conjugation $\chi_i(\tau;0,0)=\chi_{N-i}(\tau;0,0)$. On the code side, this means that the operation mapping a codeword $(a,b)$ to $(a,-b)$ becomes a code equivalence. This reduces the number of inequivalent codes, or distinct partition functions, to $\lceil\kappa_N/2\rceil$, leading to the enumeration of \cite{Meruliya:2021utr}. For the general characters we consider, the operation $(a,b)\to (a,-b)$ is not a code equivalence.
 
 \subsection{Ensemble average at $n=1$}
  
We begin by calculating the Poincar\'e series of the $su(N)_1\times \bar{su}(N)_1$ vacuum character, which can be expressed as follows
 \be \hat W=\sum_{\gamma\in SL(2,\mathbb Z)/\Gamma}\gamma(x_0\bar  x_0)\label{poincareseriesSUN},\ee
 where $\Gamma$ is a congruence subgroup of $SL(2,\mathbb Z)$ that fixes $x_0\bar x_0$. A subgroup that achieves this is
 \be \Gamma=\begin{cases}
 	\Gamma_0(N) & N\text{ odd}\\
 	\Gamma_0(2N) & N\text{ even},
 \end{cases}\ee
 where
 \be \Gamma_0(m)=\left\{\begin{pmatrix}
 	a & b\\
 	c & d
 \end{pmatrix}\in SL(2,\mathbb Z): c=0\mod m\right\}.\ee
 Its index is finite
 \be [SL(2,\mathbb Z):\Gamma_0(m)]=N\prod_{p|m,~p\text{ prime}}\left(1+{1\over p}\right) ,\label{indexGamma0}\ee
 where the product is over the prime divisors of $m$.
 
 \paragraph{\underline{Handlebody topologies:}}
 For odd prime $N=p$, the Poincar\'e series leads to
 \be \hat W=(1+\sum_{k=0}^{p-1}T^kS)x_0\bar x_0=x_0\bar x_0+\sum_{a,b\in \mathbb Z_N} x_a\bar x_b\delta_{a^2,b^2}=W_0+ W_1=2\langle W\rangle,\label{poincareprime}\ee
 while for $N=2$ we have
 \be \hat W=(1+\sum_{k=0}^3 T^k S+ST^2S)x_0\bar x_0=3 W_0=3\langle W\rangle\label{poincareprime2}\ee
 and for $N=4$
 \be \hat W=(\sum_{k=0}^7 T^k S+\sum_{k=0}^3 ST^{2k}S)x_0\bar x_0=2(W_0+W_2)=4\langle W\rangle.\label{n4poin}\ee
 Using (\ref{CRT}) this result generalizes straightforwardly to $N=2^f\prod_i {p_i}$, where $f=0,1,2$ and $p_i$ are distinct odd primes
 \be \hat W\propto {1\over \mathcal N}\sum_{\delta|^*N}W_\delta=\langle W\rangle, \ee
 where $W_\delta$ is defined in \eqref{Wd} and $\mathcal N$ is a normalization constant. By $\delta|^*N$ we denote the divisors of $N$ which do not contain a single factor of $2$.
 Therefore, for odd square-free $N$, or even $N$ such that $N/2$ is square-free, the average boundary CFT partition function is proportional to the sum over handlebody topologies.
 
 \paragraph{\underline{Contributions from singular topologies:}}
  \begin{figure}[h!]
 	\centering
 	\includegraphics[width=0.8\textwidth]{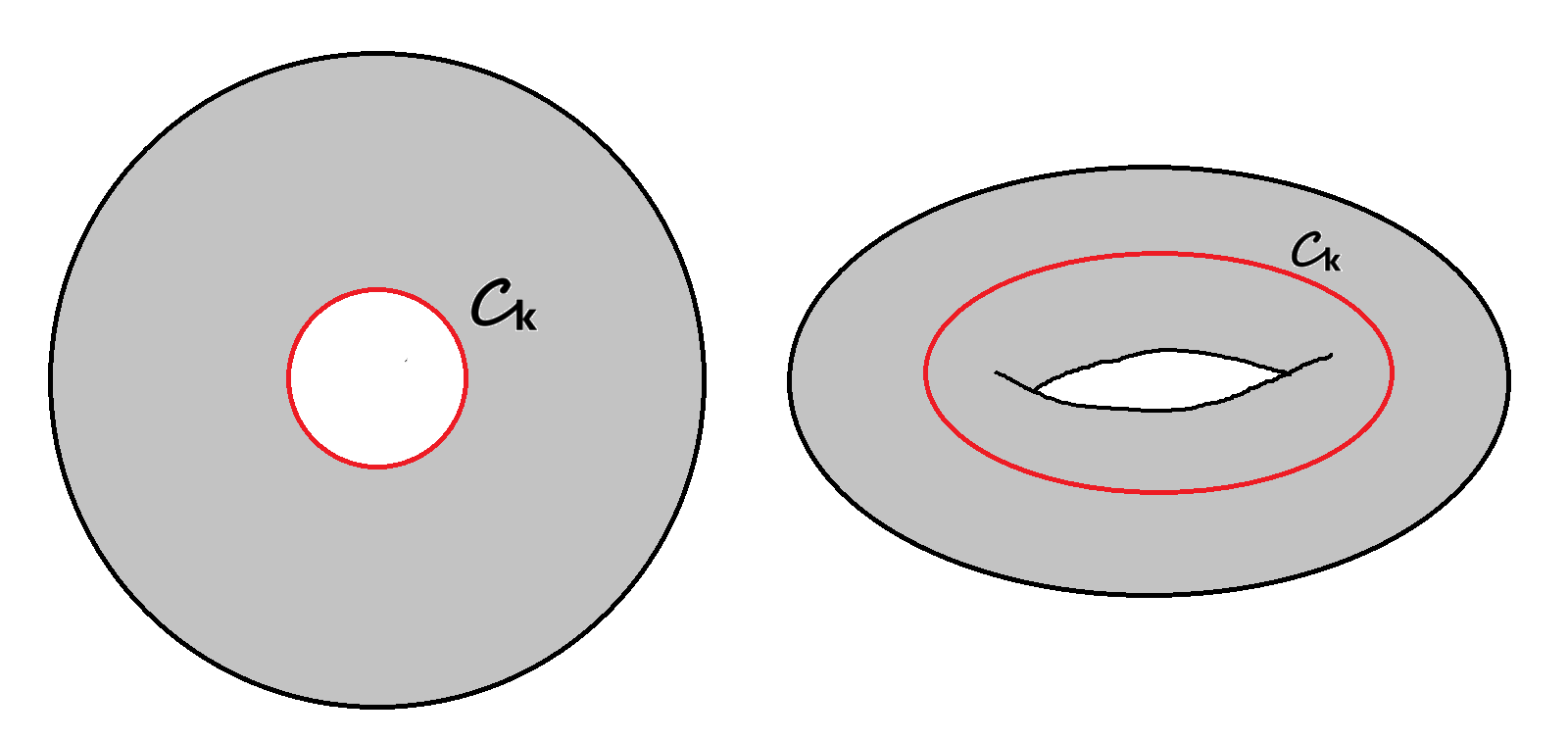}
 	\caption{\small {Left side: the annulus describing a cross-section of $T^2\times [0,1]$. The red circle represents the inner boundary, where topological boundary conditions corresponding to the code $\mathcal C_k$ have been imposed. Right side: after the the surface of the inner torus boundary has been shrunk, the resulting topology is a handlebody with a line defect.}\label{torus}}
 \end{figure}
 
For $N=p^m$, an odd prime power, the index of the congruence subgroup $\Gamma_0(p^{m})$ is 
 \be [SL(2,\mathbb Z):\Gamma_0(p^{m})]=p^{m}+p^{m-1},\ee
 and we can choose the following representatives to perform the sum
 \be \hat W=\left(\sum_{k=0}^{p^{m}-1}T^kS+\sum_{k=0}^{p^{m-1}-1}ST^{pk}S\right)x_0 \bar x_0=W_{\mathcal C_0}+W_{\mathcal C_m}+{p-1\over p}\sum_{k=1}^{m-1}W_{\mathcal C_k} .\ee
  Meanwhile, for $N=2^m$, with $m\geq 3$, we find
 \be \hat W=\left(\sum_{k=0}^{2^{m+1}-1}T^kS+\sum_{k=0}^{2^{m}-1}ST^{2k}S\right)x_0 \bar x_0=2\left(W_{\mathcal C_0}+W_{{\mathcal C_m}}+{1\over 2}\sum_{\substack{k=1\\k\neq \lfloor {m\over 2}\rfloor}}^{m-1} W_{\mathcal C_k}\right) .\ee
In either of these cases, the result is not proportional to the average partition function $\langle W\rangle$.
 For odd $N=p^m$, we can rearrange the result
 \be \langle W\rangle\propto\hat W+ {1\over p}\sum_{k=1}^{m-1} W_{\mathcal C_k}.\label{avgp}\ee

The additional terms that appear on the right-hand-side correspond to singular topologies $M_k$, $1\leq k\leq m-1$, which can be described by handlebodies, with a defect line inserted along the non-contractible cycle (see figure \ref{torus}). The presence of this defect changes the first homology group to $ H_1(M_k,\mathbb Z_{p^m})=\mathbb Z_{p^k}\times \mathbb Z_{p^{m-k}}$, since it affects which Wilson lines are contractible.

Finally, we briefly comment on the interpretation of the codes in the ``unfolded" description (where the $U(1)^{r}$ CS theory is defined on $T^2\times [0,2]$, with a surface operator \cite{KS1,KS2,Roumpedakis:2022aik} inserted at $T^2\times\{1\}$). The codes $\mathcal C_0$ and $\mathcal C_m$ define invertible surface operators (with $\mathcal C_0$ corresponding to the trivial operator). The codes $\mathcal C_1,\dots,\mathcal C_{m-1}$ define non-invertible surface operators, which modify the first homology group if inserted in an empty handlebody. Therefore, we notice that the singular contributions in \eqref{avgp} are linked to the presence of non-invertible surface operators.

\subsection{Classification of codes of length $n=2$ and prime $N=p$}

We now describe the classification of codes of length $n=2$ and prime $N=p$. The ensemble average and its holographic description is part of subsection \ref{poinsun}. 

In some cases, the generator matrix of a code can be brought into the following form by performing Gauss-Jordan elimination on its left $2\times 2$ block
\be (I|B),~~~B\in O(2,\mathbb Z_N)\ee
where $I$ is the $2\times 2$ identity matrix and $B$ is an orthogonal $2\times 2$ matrix with entries in the ring $\mathbb Z_N$. We call these codes \textit{$B$-form codes}.

\paragraph{\underline{$p=2$:}}
There are $2$ $B$-form codes, with their $B$ matrix given by
\be B\in\left\{\begin{pmatrix}
	1 & 0\\
	0 & 1
\end{pmatrix},~~\begin{pmatrix}
	0 & 1\\
	1 & 0
\end{pmatrix}\right\}.\ee
Both of these codes have the same enumerator polynomial $(x_0\bar x_0+x_1\bar x_1)^2$, which is simply the square of the $n=1$ polynomial.
\paragraph{\underline{$p=3\mod 4$:}}
 When $p=3\mod 4$, the equation $x^2+y^2=0$ has no non-trivial solutions. This implies that all codes are $B$-form codes. For these values of $p$, the orthogonal group has order $|O(2,\mathbb Z_p)|=2(p+1)$, hence there are $2(p+1)$ self-dual codes. We list generating sets for the $B$ matrices at $p=3,7,11$
\begin{eqnarray}
	p=3&:&~~B\in\left\langle\begin{pmatrix}
		 1 & 0\\
		0 & - 1
	\end{pmatrix},~~\begin{pmatrix}
		0& -1\\
		 1& 0\end{pmatrix}\right\rangle
		\\
	 p=7&:&~~B\in\left\langle\begin{pmatrix}
	 	1 & 0\\
	 	0 & - 1
	 \end{pmatrix},~~\begin{pmatrix}
	 	0& -1\\
	 	1& 0\end{pmatrix},~~\begin{pmatrix}
		2& 2\\
		 2&  -2
	\end{pmatrix}\right\rangle\\
	p=11&:&~~B\in\left\langle\begin{pmatrix}
		1 & 0\\
		0 & - 1
	\end{pmatrix},~~\begin{pmatrix}
		0& -1\\
		1& 0\end{pmatrix},~~\begin{pmatrix}
		3&5\\
		5& -3
	\end{pmatrix}\right\rangle.
\end{eqnarray}

\paragraph{\underline{$p=1\mod 4$:}}
In this case, the order of the orthogonal group is $|O(2,\mathbb Z_p)|=2(p-1)$, hence there are $2(p-1)$ $B$-form codes.
Generating sets for the $B$ matrices at $p=5,13,17$ are given by
\begin{eqnarray}
	p=5&:&~~B\in\left\langle\begin{pmatrix}
		1 & 0\\
		0 & - 1
	\end{pmatrix},~~\begin{pmatrix}
		0& -1\\
		1& 0\end{pmatrix}\right\rangle
	\\
	p=13&:&~~B\in\left\langle\begin{pmatrix}
		1 & 0\\
		0 & - 1
	\end{pmatrix},~~\begin{pmatrix}
		0& -1\\
		1& 0\end{pmatrix},~~\begin{pmatrix}
		2&6\\
		6& - 2
	\end{pmatrix}\right\rangle.
	\\
	p=17&:&~~B\in\left\langle\begin{pmatrix}
		1 & 0\\
		0 & - 1
	\end{pmatrix},~~\begin{pmatrix}
		0& -1\\
		1& 0\end{pmatrix},~~\begin{pmatrix}
		3&3\\
		3& - 3
	\end{pmatrix},~~\begin{pmatrix}
		4&6\\
		6& - 4
	\end{pmatrix}\right\rangle.
\end{eqnarray}
 In addition, there are $4$ codes, which are not $B$-form, bringing the total up to $2(p+1)$. They are direct sums of Euclidean self-dual codes of length $2$ and their generators are given by
\be \left(\begin{array}{cc|cc}
	1 & \pm i & 0 & 0\\
	0 & 0 & 1 & \pm i
\end{array}\right) ,\ee
where $i\in\mathbb Z_p$ is such that $i^2=-1$. The construction A lattice (\ref{consA}) of these codes is a direct sum of two even, self-dual Euclidean lattices. The dimension of these lattices is $2(p-1)$, which is a multiple of $8$. These codes describe non-invertible surface operators in the ``unfolded" theory, however they do not correspond to singular topologies.

\subsection{Ensemble average at $n=2$ for $N=4$}\label{sun4}

In this case there are 8 $B$-form codes, with $B$ matrices given by
\be B\in\left\{\begin{pmatrix}
	\pm 1 & 0\\
	0 & \pm 1
\end{pmatrix},~~\begin{pmatrix}
	0&  \pm 1\\
	\pm 1 & 0 
\end{pmatrix}\right\}.\label{Bform}\ee
Among these $8$ codes, there are 3 distinct enumerator polynomials, which belong to the polynomial ring generated by the $2$ invariants at $n=1$:
\be w_1=\sum_{i=0}^3 x_i\bar x_i,~~w_2=\sum_{i=0}^3 x_i\bar x_{-i}.\ee
The distinct polynomials corresponding to (\ref{Bform}) can be written as $w_1^2,w_2^2,w_1w_2$ with multiplicities $2,2,4$ respectively.

There are $2$ additional even, self-dual code with generator matrices
\be C_9=\left(\begin{array}{cc|cc}
	1 &  1 & 1 & 1\\
	0 & 2 & 0 & 2\\
	0 & 0 & 2 & 2
\end{array}\right),~~C_{10}=\left(\begin{array}{cc|cc}
1 &  3 & 1 & 1\\
0 & 2 & 0 & 2\\
0 & 0 & 2 & 2
\end{array}\right) ,\label{W0}\ee
whose enumerator polynomials are
\be W_9=(x_0\bar x_0+x_2\bar x_2)^2+(x_0\bar x_2+x_2\bar x_0)^2+(x_1\bar x_1+x_3\bar x_3)^2+(x_1\bar x_3+x_3\bar x_1)^2,\ee
\be W_{10}=(x_0\bar x_0+x_2\bar x_2)^2+(x_0\bar x_2+x_2\bar x_0)^2+2(x_3\bar x_1+x_1\bar x_3)(x_1\bar x_1+x_3\bar x_3).\ee
Let us now evaluate the Poincar\'e sum as in (\ref{n4poin})
\be \hat W=(\sum_{k=0}^7 T^k S+\sum_{k=0}^3 ST^{2k}S)x_0^2\bar x_0^2={1\over 2}(w_1+w_2)^2 .\ee
We see that this sum is proportional to the average over $B$-form codes only. The codes $C_9$, $C_{10}$ do not appear. This is a genus $1$ effect, meaning that at larger genus, all codes do appear in the Poincar\'e series of the vacuum character.  The average enumerator polynomial can be expressed in terms of $\hat W$ as follows
\be \langle W\rangle\propto \hat W+{1\over 4}(W_9+W_{10}).\ee
Unlike in the $n=1$ case, we now have additional contributions beyond the sum over handlebodies.  Similarly to (\ref{avgp}), this is due to the presence of  non-invertible surface operators in the ``unfolded" theory, described by the codes $C_9$ and $C_{10}$.

\subsection{Ensemble average at arbitrary $n$ for prime $N=p$}\label{poinsun}
We now compute the average code enumerator polynomial for prime $N=p$ and any code length $n$. Since codes related by permutations of their first or last $n$ letters yield the same enumerator polynomial, not all polynomials are distinct.
When calculating the average enumerator polynomial, we must take an equal-weighted average over \textit{all} codes, rather than only over those with distinct polynomials. We will demonstrate that this average coincides with the Poincaré series of the vacuum for all $n$ and prime $N=p$. By application of the CRT (\ref{CRT}), this equality extends to the case when $N$ is square-free. Due to subtle differences in the case $p=2$, we treat it separately.

To compute the average enumerator polynomial, we closely follow the method outlined in \cite{Kawabata:2022jxt}.

\paragraph{\underline{$N=p>2$:}}

To each element $(\vec a,\vec b)\in G^n\times\bar G^n$ we assign a pair of tuples
\be A= (e_0(\vec a),e_1(\vec a),\dots, e_{p-1}(\vec a)),~~\bar A=(e_0(\vec b),e_1(\vec b),\dots,e_{p-1}(\vec b)).\ee
where $e_g(\vec a)$, as defined in (\ref{eg}), counts the entries of $\vec a$ equal to $g$. A pair of tuples $A,\bar A$ is called \textit{admissible}, if and only if the corresponding codeword belongs to a self-orthogonal code of length $n$. Hence, admissible pairs must satisfy
\be \sum_{i=0}^{p-1} A_i=\sum_{i=0}^{p-1}\bar A_i=n,~~~\sum_{i=0}^{p-1}(A_i-\bar A_i) i^2=0\mod p.\label{admissible}\ee
Let $\mathcal N_{\vec a,\vec b}$ denote the number of self-dual codes that contain the codeword $(\vec a,\vec b)$. It is given by
\be \mathcal N_{\vec a,\vec b}=\begin{cases}
	\prod_{i=0}^{n-2}(p^i+1) & (\vec a,\vec b)\neq (\vec 0,\vec 0)\land \langle (\vec a,\vec b)|(\vec a,\vec b)\rangle=0\\
	\prod_{i=0}^{n-1}(p^i+1) & (\vec a,\vec b)=(\vec 0,\vec 0)\\
	0&\text{ otherwise}.
\end{cases}\ee
Note that the total number of self-dual codes is $\mathcal N_{\vec 0,\vec 0}$.

The average enumerator polynomial is, by definition
\be 	\langle W\rangle\equiv{1\over \mathcal N_{\vec 0,\vec 0}}\sum_{\text{self-dual }C}W_C(\{x_i,\bar x_i\})={1\over \mathcal N_{\vec 0,\vec 0}}\sum_{\text{self-dual }C}\sum_{(\vec a,\vec b)\in C}x^{A(\vec a)}\bar x^{\bar A(\vec b)},\label{avgpol}\ee
where we used the shorthand notation
\be x^{A(\vec a)}=x_0^{e_0(\vec a)}\dots x_{p-1}^{e_{p-1}(\vec a)},~~\bar x^{\bar A(\vec b)}=\bar x_0^{e_0(\vec b)}\dots \bar x_{p-1}^{e_{p-1}(\vec b)}.\ee
The sum on the RHS of (\ref{avgpol}) can equivalently be expressed as a sum over all codewords  $(\vec a,\vec b)\in G^n\times\bar G^n$, weighted by the number of self-dual codes in which $(\vec a,\vec b)$ appears
\be \langle W\rangle={1\over \mathcal N_{\vec 0,\vec 0}}\sum_{(\vec a,\vec b)\in G^n\times \bar G^n}\mathcal N_{\vec a,\vec b}x^{A(\vec a)}\bar x^{\bar A(\vec b)}=x_0^n\bar x_0^n+{1\over 1+p^{n-1}}\sum_{\substack{\text{admissible }A,\bar A\\ A_0+\bar A_0<2n}}\binom{n}{A}\binom{n}{\bar A}x^A \bar x^{\bar A}.\label{averg}\ee
At the last step we isolated the contribution from the zero codeword and rewrote the sum in terms of the remaining admissible tuples $A,\bar A$ (\ref{admissible}). We also introduced the combinatorial factors
\be \binom{n}{A}\equiv {n!\over A_0!A_1!\dots A_{p-1}!},\ee
which appear because codewords related by permutations of the first $n$ or last $n$ coordinates give rise to the same monomial.

We now turn to the calculation of the Poincar\'e series. The stabilizer of the vacuum character $x_{\vec 0}\bar x_{\vec 0}$ is $\Gamma_0(p)$ and we can choose the same representatives as in (\ref{poincareprime})
\be \hat W=(1+\sum_{k=0}^{p-1}{T^kS})x_0^n\bar x_0^n=x_0^n\bar x_0^n+{1\over p^n}\sum_{r=0}^{p-1}\sum_{(\vec a,\vec b)\in G^n\times\bar G^n}x_{\vec a}\bar x_{\vec b}e^{-{(p-1)\pi i\over p}(\vec a^2-\vec b^2)r}.\ee
Writing $x_{\vec a}\bar x_{\vec b}=x^{A(\vec a)}\bar x^{\bar A(\vec b)}$ and converting the sum over codewords into a sum over tuples $A,\bar A$, we obtain symmetry factors $\binom{n}{A}\binom{n}{\bar A}$
\be \hat W=x_0^n\bar x_0^n+{1\over p^{n-1}}x_0^n\bar x_0^n+{1\over p^{n}}\sum_{r=0}^{p-1}\sum_{\substack{A,\bar A\\ A_0+\bar A_0<2n}}x^A\bar x^{\bar A}\binom{n}{A}\binom{n}{\bar A}e^{-{(p-1)\pi i\over p}r\sum_{j=0}^{p-1}(A_j-\bar A_j)j^2}.\ee
The sum over $r$ enforces the condition on the right side of (\ref{admissible}). Comparing with (\ref{averg}) we obtain
\be \hat W={p^{n-1}+1\over p^{n-1}}\langle W\rangle\propto \langle W\rangle.\ee

\paragraph{\underline{${N=p=2}$:}}

The $p=2$ case has a few subtle differences. First, the evenness condition is not satisfied by all self-dual codes. An admissible pair of tuples $A,\bar A$ satisfies
\be A_0+\bar A_1=\bar A_0+\bar A_1=n,~~~A_1-\bar A_1=0\mod 4.\label{2admissible}\ee
Additionally, every binary even, self-dual code contains the codeword with all entries equal to $1$, which we denote by $(\vec 1,\vec 1)$. This modifies the counting of the even, self-dual codes containing a codeword $(\vec a,\vec b)$
\be \mathcal N_{\vec a,\vec b}=\begin{cases}
	\prod_{i=1}^{n-2}(2^{i-1}+1) & (\vec a,\vec b)\neq (\vec 0,\vec 0)\land (\vec a,\vec b)\neq (\vec 1,\vec 1)\land wt(\vec a)-wt(\vec b)=0\mod 2\\
	\prod_{i=1}^{n-1}(2^{i-1}+1) & (\vec a,\vec b)=(\vec 0,\vec 0)\lor (\vec a,\vec b)= (\vec 1,\vec 1)\\
	0&\text{ otherwise}.
\end{cases}\label{p2count}\ee

We calculate the average polynomial similarly to $p>2$, but now we also isolate the monomial $x_1^n\bar x_1^n$
\be \langle W\rangle=x_0^n\bar x_0^n+x_1^n\bar x_1^n+{1\over 1+2^{n-2}}\sum_{\substack{\text{admissible }A,\bar A\\ 1<A_0+\bar A_0<2n}}\binom{n}{A}\binom{n}{\bar A}x^A \bar x^{\bar A}.\label{averg2}\ee
The Poincar\'e series can be calculated as in (\ref{poincareprime2})
\be \hat W=(1+ST^2S+\sum_{k=0}^3 T^k S)x_0^n\bar x_0^n=x_0^n\bar x_0^n+x_1^n\bar x_1^n+{1\over 2^n}\sum_{r=0}^3 \sum_{A,\bar A}\binom{n}{A}\binom{n}{\bar A}x^A\bar x^{\bar A} e^{{\pi i r\over 2}(A_1-\bar A_1)},\ee
where we used that $ST^2S(x_0^n\bar x_0^n)=x_1^n\bar x_1^n$. Comparing with (\ref{averg2}) we find
\be \langle W\rangle\propto \hat W.\ee

\paragraph{\underline{square-free $N$:}}

Applying the map (\ref{CRT}) to the ring $\mathbb Z_N$ when $N$ is square-free yields a product of rings of prime order. The averages and Poincar\'e sums can be evaluated independently over each factor and then combined using the inverse map $\pi^{-1}$. Consequently, since the equality  between the average and the Poincaré series holds for $\mathbb Z_p$ for all primes $p$, it follows that the same equality holds for $\mathbb Z_N$ when $N$ is square-free.

 \section{The $so(2r)^n_1$ WZW models}\label{sec:dn}
 
 The $so(2r)_1$ ($r\geq 4$) algebra has rank $r$ and 4 simple roots of comark 1 (independent of $r$), hence there are always $4$ dominant weights at level $1$.
 The root lattice of $so(2r)$ is $\Lambda=D_r$, with a generator matrix given by
 \be \Lambda_{ij}=\begin{cases}
 	1 & i=j\\
 	-1 & i=j+1\\
 	1 & i=r-1\land j=r\\
 	0 & \text{otherwise}
 \end{cases}.\ee
 The discriminant group depends on the parity of $r$
 \be G=\Lambda^\perp/\Lambda\cong\begin{cases}
 	\mathbb Z_4 & r\text{ odd}\\
 	\mathbb Z_2\times \mathbb Z_2 & r\text{ even}.
 \end{cases}\ee

 \subsection{Modular invariant CFTs and ensemble average at $n=1$ for odd $r$}

 For odd $r$, $G$ is cyclic. The element $\lambda=({1\over2},{1\over 2},\dots,{1\over 2})^T\in\Lambda^\perp$ has order $4$ in $\Lambda^\perp/\Lambda$ and we choose the map $\phi$ such that $\phi(\lambda)=1\in\mathbb Z_4$. 
 This leads to the bilinear form on $G$
 \be \langle a_1|a_2\rangle={r\over 4}a_1a_2\mod \mathbb Z \label{ipG}\ee
 and the weights
 \be wt(0)=0,~wt(1)={r\over 4},~wt(2)={1},~wt(3)={r\over 4} .\ee
 The characters can be written in terms of Jacobi theta functions
 \bea
 \chi_0(\tau;\xi,t)&=&{e^{-2\pi i t}\over 2\eta^r}\left(\prod_{i=1}^r\theta_3(\pi \xi_i,q)+\prod_{i=1}^r\theta_4(\pi\xi_i,q)\right)\\
 \chi_1(\tau;\xi,t)&=&{e^{-2\pi i t}\over 2\eta^r}\left(\prod_{i=1}^r\theta_2(\pi\xi_i,q)-i^r \prod_{i=1}^r\theta_1(\pi\xi_i,q)\right)\\
 \chi_2(\tau;\xi,t)&=&{e^{-2\pi i t}\over 2\eta^r}\left(\prod_{i=1}^r\theta_3(\pi\xi_i,q)-\prod_{i=1}^r\theta_4(\pi\xi_i,q)\right)\\
 \chi_3(\tau;\xi,t)&=&{e^{-2\pi i t}\over 2\eta^r}\left(\prod_{i=1}^r\theta_2(\pi\xi_i,q)+i^r \prod_{i=1}^r\theta_1(\pi\xi_i,q)\right).
 \eea
 The $T,S$ matrices are, explicitly
 \be T_{aa'}=e^{\pi i {a^2r\over 4}}\delta_{aa'} ,~~S_{aa'}={1\over 2}e^{-2\pi i {raa'\over 4}}.\ee
There exist $2$ even, self-dual codes, with generators 
 \be C_0= (11),~C_1= (13),\ee
 leading to the CFT partition functions
 \be W_0=\sum_{i=0}^3 x_i\bar x_i,~W_1=\sum_{i=0}^3 x_i\bar x_{-i}.\ee
 An (infinite) subgroup of $SL(2,\mathbb Z)$ that leaves the vacuum character invariant is $\Gamma_0(8)$, of index $12$.
 
 Let us consider the holographic interpretation of this ensemble. By calculating the sum over modular images of the vacuum
 \be \hat W=\sum_{\gamma\in SL(2,\mathbb Z)/\Gamma_0(8)}\gamma(x_0\bar  x_0)=\left(\sum_{k=0}^{7}T^kS+\sum_{k=0}^{3}ST^{2k}S\right)x^i_0 \bar x^i_0=W_0+W_1=2\langle W\rangle\label{poincareseriescodeDn}.\ee
we find that the average partition function is proportional to sum over handlebodies
 \be \langle W\rangle={1\over 2}(W_0+W_1)\propto\hat W.\ee
 
  \subsection{Modular invariant CFTs and ensemble average at $n=1$ for even $r$}
  
   For even $r$, $G=\mathbb Z_2\times\mathbb Z_2$, therefore we must find $2$ generators. We choose the vectors $\lambda_\pm=({1\over 2},{1\over 2},\dots,\pm{1\over 2})^T\in \Lambda^\perp$  and the map $\phi$ such that $\phi(\lambda_+)=(1,0)$ and $\phi(\lambda_-)=(0,1)$.
  This leads to the bilinear form on $G$
  \be \langle (a,b)|(a',b')\rangle={r\over 4}(aa'+bb')+{r-2\over 4}(ab'+a'b)\mod \mathbb Z\label{bf1} ,\ee
  and the weights
  \be wt(0,0)=0,~wt(1,1)=1,~wt(1,0)=wt(0,1)={r\over 4}.\label{wts}\ee
  The  holomorphic characters can be written in terms of Jacobi theta functions, as in the case of odd $r$
 \bea
\chi_{(0,0)}(\tau;\xi,t)&=&{e^{-2\pi i t}\over 2\eta^r}\left(\prod_{i=1}^r\theta_3(\pi \xi_i,q)+\prod_{i=1}^r\theta_4(\pi\xi_i,q)\right)\\
\chi_{(1,0)}(\tau;\xi,t)&=&{e^{-2\pi i t}\over 2\eta^r}\left(\prod_{i=1}^r\theta_2(\pi\xi_i,q)-i^r \prod_{i=1}^r\theta_1(\pi\xi_i,q)\right)\\
\chi_{(1,1)}(\tau;\xi,t)&=&{e^{-2\pi i t}\over 2\eta^r}\left(\prod_{i=1}^r\theta_3(\pi\xi_i,q)-\prod_{i=1}^r\theta_4(\pi\xi_i,q)\right)\\
\chi_{(0,1)}(\tau;\xi,t)&=&{e^{-2\pi i t}\over 2\eta^r}\left(\prod_{i=1}^r\theta_2(\pi\xi_i,q)+i^r \prod_{i=1}^r\theta_1(\pi\xi_i,q)\right).
\eea
  The $T,S$ matrices are
  \be T_{(a,b),(a',b')}=e^{\pi i ((a^2+b^2){r\over 4}+ab{r-2\over 2})}\delta_{aa'}\delta_{bb'} ,\ee
  \be S_{(a,b),(a',b')}={1\over 2}e^{-2\pi i ({r\over 4}(aa'+bb')+{r-2\over 4}(ab'+a'b))}.\ee
  
  Due to the form of \eqref{bf1}, the classification of even, self-dual codes depends on the residue of $r$ modulo $8$. Therefore, we treat each case separately.
 
 \subsubsection{$r=2\mod 4$}
 
 In this case the bilinear form (\ref{bf1}) simplifies to
 \be \langle (a,b)|(a',b')\rangle={1\over 2}(aa'+bb')\mod\mathbb Z.\ee
 There are two even, self-dual codes, generated by the rows of the following matrices
 \be C_0=\begin{pmatrix}
 	(10)&(10)\\(01)&(01)\end{pmatrix},~C_1=  \begin{pmatrix}(01)&(10)\\(10)&(01)\end{pmatrix},\ee
 giving rise to the diagonal and conjugation-symmetric modular invariants respectively
 \be W_0=x_{(00)}\bar x_{(00)}+x_{(10)}\bar x_{(10)}+x_{(01)}\bar x_{(01)}+x_{(11)}\bar x_{(11)},\ee
 \be W_1=x_{(00)}\bar x_{(00)}+x_{(10)}\bar x_{(01)}+x_{(01)}\bar x_{(10)}+x_{(11)}\bar x_{(11)}.\ee
 
 We now consider the Poincar\'e series of the vacuum character. A  subgroup of $SL(2,\mathbb Z)$ that leaves the vacuum character invariant is $\Gamma_0(4)$ and we find
 \be \hat W=(\sum_{i=0}^{3}T^iS+\sum_{i=0}^1 ST^{2i}S)x_{(00)}\bar x_{(00)} =W_1+W_2.\label{req2mod4}\ee 
Again, the average CFT partition function is proportional to the sum over handlebodies
 \be \langle W\rangle \propto W_0+W_1=\hat W.\ee
 
 \subsubsection{$r=4\mod 8$}
In this case the weights (\ref{wts}) are odd integers, while the bilinear form (\ref{bf1}) reads
 \be \langle (a,b)|(a',b')\rangle={1\over 2}(ab'+a'b)\mod\mathbb Z .\ee
 There are six even, self-dual codes
 \be C_0=\begin{pmatrix}
 	(10)&(10)\\(01)&(01)\end{pmatrix},~C_1=  \begin{pmatrix}(01)&(10)\\(10)&(01)\end{pmatrix},\ee
 \be C_2=\begin{pmatrix}
 	(10)&(11)\\(11)&(10)\end{pmatrix},~C_3=  \begin{pmatrix}(01)&(11)\\(11)&(10)\end{pmatrix},\ee
 \be C_4=\begin{pmatrix}
 	(10)&(11)\\(11)&(01)\end{pmatrix},~C_5=  \begin{pmatrix}(01)&(11)\\(11)&(01)\end{pmatrix},\ee
 giving rise to the modular invariants
 \bea
   W_0&=& x_{(00)}\bar x_{(00)}+ x_{(10)}\bar x_{(10)}+ x_{(01)}\bar x_{(01)}+ x_{(11)}\bar x_{(11)},\\
  W_1&=& x_{(00)}\bar x_{(00)}+ x_{(10)}\bar x_{(01)}+ x_{(01)}\bar x_{(10)}+ x_{(11)}\bar x_{(11)},\\
  W_2&=& x_{(00)}\bar x_{(00)}+ x_{(10)}\bar x_{(11)}+ x_{(11)}\bar x_{(10)}+ x_{(01)}\bar x_{(01)},\\
  W_3&=& x_{(00)}\bar x_{(00)}+ x_{(01)}\bar x_{(11)}+ x_{(11)}\bar x_{(10)}+ x_{(10)}\bar x_{(01)},\\
  W_4&=& x_{(00)}\bar x_{(00)}+ x_{(10)}\bar x_{(11)}+ x_{(11)}\bar x_{(01)}+ x_{(01)}\bar x_{(10)},\\
  W_5&=& x_{(00)}\bar x_{(00)}+ x_{(01)}\bar x_{(11)}+ x_{(11)}\bar x_{(01)}+ x_{(10)}\bar x_{(10)}.
 \eea
 Since the group $\mathbb Z_2\times\mathbb Z_2$ has 5 subgroups, the orbifolding procedure in appendix \ref{appOrbifold} leads to 5, rather than 6, modular invariants. This is because even with flavored characters the above CFT partition functions are not linearly independent, but satisfy
 \be W_0+W_3+W_4=W_1+W_2+W_5.\label{linconst}\ee

 We now calculate the Poincar\'e series of the vacuum character. The $S,T$ matrices in this representation satisfy the relations of the dihedral group $D_6$: $S^2=T^2=(ST)^3=1$, which is a finite group of order $6$. Moreover, $T$ stabilizes the vacuum character. Therefore the series has only $3$ terms
 \be \hat W=(1+S+TS)x_{(00)}\bar x_{(00)} ={1\over 2}(W_0+W_3+W_4).\ee
It may seem that the Poincar\'e series results in an average over half of the modular invariants,  but due to (\ref{linconst}) there is an ambiguity in interpreting this result.  Using \eqref{linconst} we can write $\langle W\rangle=2(W_0+W_3+W_4)$, therefore the average CFT partition function can again be expressed as a sum over handlebodies
 \be \langle  W\rangle\propto\hat W.\ee
 
 \subsubsection{$r=0\mod 8$}
  In this case the weights (\ref{wts}) of $(1,0)$ and $(0,1)$ are even integers, and the bilinear form (\ref{bf1}) simplifies to
 \be \langle (a,b)|(a',b')\rangle={1\over 2}(ab'+ba')\mod\mathbb Z.\ee
 There are six even, self-dual codes
 \be C_0=\begin{pmatrix}
 	(10)&(10)\\(01)&(01)\end{pmatrix},~C_1=  \begin{pmatrix}(01)&(10)\\(10)&(01)\end{pmatrix},\ee
 \be C_2=\begin{pmatrix}
 	(01)&(00)\\(00)&(01)\end{pmatrix},~C_3=  \begin{pmatrix}(01)&(00)\\(00)&(10)\end{pmatrix},\ee
 \be C_4=\begin{pmatrix}
 	(10)&(00)\\(00)&(01)\end{pmatrix},~C_5=  \begin{pmatrix}(10)&(00)\\(00)&(10)\end{pmatrix},\ee
 giving rise to the modular invariants
 \bea 
  W_0&=& x_{00}\bar x_{00}+ x_{10}\bar x_{10}+ x_{01}\bar x_{01}+ x_{11}\bar x_{11},\\
  W_1&=& x_{00}\bar x_{00}+ x_{01}\bar x_{10}+ x_{10}\bar x_{01}+ x_{11}\bar x_{11},\\
  W_2&=&( x_{00}+ x_{01})( \bar x_{00}+ \bar x_{01}),\\
  W_3&=&( x_{00}+ x_{01})(\bar x_{00}+\bar x_{10}),\\
  W_4&=&( x_{00}+ x_{10})(\bar x_{00}+\bar x_{01}),\\
  W_5&=&( x_{00}+ x_{10})(\bar x_{00}+\bar x_{10}).
 \eea
As in the previous case, the CFT partition functions are linearly dependent
\be W_0+W_3+W_4=W_1+W_2+W_5.\ee
 
 We calculate the Poincar\'e series of the vacuum character similarly to the previous case:
 \be \hat W=(1+S+TS)x_{(00)}\bar x_{(00)} ={1\over 2} (W_0+W_3+W_4),\ee 
and writing $\langle W\rangle=2(W_0+W_3+W_4)$ we obtain
 \be \langle W\rangle\propto\hat W.\ee
 
 \subsection{Ensemble average at $n=2$ for odd $r$}
 
 For odd $r$, the discriminant group is $G=\mathbb Z_4$. The analysis is very similar to subsection \ref{sun4}, since the bilinear form and evenness conditions are the same. We find $10$ even, self-dual codes.
 There are 8 $B$-form codes given by (\ref{Bform}) and 2 additional codes $C_9, C_{10}$ given by (\ref{W0}). We emphasize that even though the codes are the same as in section \ref{sun4}, the CFT partition functions they give rise to are different, due to the dependence of construction A (\ref{consA}) on the root lattice.
The average enumerator polynomial can be expressed in terms of the Poincar\'e series $\hat W$ as follows
\be \langle W\rangle\propto \hat W+{1\over 4}(W_9+W_{10}),\ee
where we find additional contributions, as in \ref{sun4}.

 
 
 
 

     
\subsection{Ensemble average for arbitrary code length $n$ and even $r$}\label{sunalln}

For even $r$, the discriminant group is $G=\mathbb Z_2\times\mathbb Z_2$. Using the group isomorphism $G^n=(\mathbb Z_2\times\mathbb Z_2)^n\cong \mathbb Z_2^{2n}$ we can view a code of length $n$ over $G$ as a binary code of length $2n$. Since the bilinear form and evenness conditions depend on the residue of $r$ modulo $8$, we treat each case separately. In all cases, we show that the average polynomial is proportional to the Poincar\'e sum of the vacuum.

 \subsubsection{$r=2\mod 4$}

Using $G^n=(\mathbb Z_2\times\mathbb Z_2)^n\cong \mathbb Z_2^{2n}$, we can express the bilinear form on $\mathbb Z^{2n}$ as follows
\be \langle \vec a|\vec a'\rangle={1\over 2}\vec a\cdot\vec a'\mod\mathbb Z.\ee 
The evenness condition on a codeword $(\vec a,\vec b)\in G^n\times\bar G^n$ is 
\be wt(\vec a)-wt({\vec b})={1\over 2}(\vec a^2-{\vec b}^2)=0\mod 2\mathbb Z.\ee
We recognize that the bilinear form and eveness condition are the same as in section (\ref{sec:an}) for $N=2$. Using the counting in (\ref{p2count}), but with $n$ replaced by $2n$, we find the average code polynomial
\be \langle W\rangle=x_{00}^n\bar x_{00}^n+x_{11}^n\bar x_{11}^n+{1\over 1+2^{2n-2}}\sum_{\substack{\text{admissible }A,\bar A\\ 1<A_0+\bar A_0<2n}}\binom{n}{A}\binom{n}{\bar A}x^A \bar x^{\bar A}.\ee
The admissible $A,\bar A$ are given by
\be A_{00}+A_{10}+A_{01}+A_{11}=\bar A_{00}+\bar A_{10}+\bar A_{01}+\bar A_{11}=n,~~A_{10}+A_{01}+2A_{11}=\bar A_{10}+\bar A_{01}+2\bar A_{11}\mod 4\mathbb Z.\ee

The Poincar\'e series is evaluated by choosing the same representatives as in (\ref{req2mod4})
\be \hat W=(1+ST^2S+\sum_{k=0}^3 T^kS)x_{00}^n\bar x_{00}^n=x_{00}^n\bar x_{00}^n+x_{11}^n\bar x_{11}^n+{1\over 4^{n-1}}\sum_{\text{admissible }A,\bar A}\binom{n}{A}\binom{n}{\bar A}x^A \bar x^{\bar A} ,\ee
leading to
\be \langle W\rangle \propto \hat W.\ee

\subsubsection{$r=0,4\mod 8$}

Finally, we treat the cases where $r$ is a multiple of $4$.
The bilinear form on $G^n\times \bar G^n$ can be written as 
\be \langle ((\vec a,\vec b),(\bar{\vec a},\bar{\vec b}))|((\vec a',\vec b'),(\bar{\vec a}',\bar{\vec b}'))\rangle={1\over 2}(\vec a',\bar {\vec a}',\vec b',\bar{\vec b}')\begin{pmatrix}
	0_{n\times n} &0_{n\times n} & I_{n\times n}&0_{n\times n} \\
	0_{n\times n} &0_{n\times n} &0_{n\times n} &I_{n\times n} &\\
	I_{n\times n} & 0_{n\times n} &0_{n\times n} &0_{n\times n} &\\
	0_{n\times n} & I_{n\times n} &0_{n\times n} &0_{n\times n} &
\end{pmatrix} \begin{pmatrix}
	\vec a^T\\
	\bar{\vec a}^T\\
	\vec b^T\\
	\bar{\vec b}^T
\end{pmatrix}\mod \mathbb Z .\ee
The codes in this class are binary quantum stabilizer codes (see \cite{Dymarsky:2020qom} for an introduction to these codes and their relation to Narain lattices).
 However, the evenness condition depends on the value of $r$.

 \paragraph{\underline{$r=4\mod 8$}} The evenness condition of a codeword $((\vec a,\vec b),(\vec a',\vec b'))\in G^n\times \bar G^n$ is the usual one for a binary quantum stabilizer code \cite{Dymarsky:2020qom}
\be wt(\vec a,\vec b)-wt(\vec a',\vec b')=\vec a^2+\vec b^2+\vec a\cdot\vec b+\vec a'^2+\vec b'^2+\vec a'\cdot\vec b'=0\mod 2\mathbb Z.\ee
This implies that the admissible tuples $A,\bar A$ satisfy
\be A_{00}+A_{10}+A_{01}+A_{11}=\bar A_{00}+\bar A_{10}+\bar A_{01}+\bar A_{11}=n,~~A_{10}+A_{01}+A_{11}=\bar A_{10}+\bar A_{01}+\bar A_{11}\mod 2\mathbb Z .\ee
  \paragraph{\underline{$r=0\mod 8$}} In this case the evenness condition of a codeword $((\vec a,\vec b),(\vec a',\vec b'))\in G^n\times \bar G^n$ is
  \be wt(\vec a,\vec b)-wt(\vec a',\vec b')=\vec a\cdot\vec b+\vec a'\cdot\vec b'=0\mod 2\mathbb Z \ee
  and the admissible tuples $A,\bar A$ satisfy
  \be A_{00}+A_{10}+A_{01}+A_{11}=\bar A_{00}+\bar A_{10}+\bar A_{01}+\bar A_{11}=n,~~A_{11}=\bar A_{11}\mod 2\mathbb Z .\ee
 
  In either case, the counting of even, self-dual codes containing the codeword $(\vec g,\vec g')\in G^n\times\bar G^n$ is
 \be \mathcal N_{\vec g,\vec g'}=\begin{cases}
 	\prod_{i=0}^{2n-2}(2^{i}+1) & (\vec g,\vec g')\neq ((\vec 0,\vec 0),(\vec 0,\vec 0))\land wt(\vec g)-wt(\vec g')=0\mod 2\\
 	\prod_{i=0}^{2n-1}(2^{i}+1) & (\vec g,\vec g')=((\vec 0,\vec 0),(\vec 0,\vec 0))\\
 	0&\text{ otherwise}.
 \end{cases}\label{dcount}\ee
 This leads to the average polynomial
 \be \langle W\rangle=x_{00}^n\bar x_{00}^n+{1\over 1+2^{2n-1}}\sum_{\substack{\text{admissible }A,\bar A\\ 1<A_0+\bar A_0<2n}}\binom{n}{A}\binom{n}{\bar A}x^A \bar x^{\bar A}.\label{d4avg}\ee
 Comparing it with the Poincar\'e series of the vacuum character
 \be \hat W=(1+S+TS)x_{00}^n\bar x_{00}^n=x_{00}^n\bar x_{00}^n+ {1\over 4^n}\sum_{r=0}^1\sum_{(\vec a,\vec b),(\vec a',\vec b')}x_{(\vec a,\vec b)}\bar x_{(\vec a',\vec b')}(-1)^{k(wt(\vec a,\vec b)-wt(\vec a',\vec b'))}\ee
 we obtain
 \be \langle W\rangle\propto \hat W.\ee

 \section{The $(E_r)^n_1$ WZW models}\label{sec:en}
 
 \subsection{$E_6$}
  
  The exceptional $E_6$ algebra has $3$ dominant weights at level $1$.
 A generator of the root lattice $\Lambda=E_6$ is
 \be \Lambda=\left(
 \begin{array}{cccccc}
 	1 & 0 & 0 & 0 & -\frac{1}{2} & 0 \\
 	-1 & 1 & 0 & 0 & -\frac{1}{2} & 0 \\
 	0 & -1 & 1 & 0 & -\frac{1}{2} & 0 \\
 	0 & 0 & -1 & 1 & -\frac{1}{2} & 1 \\
 	0 & 0 & 0 & 1 & -\frac{1}{2} & -1 \\
 	0 & 0 & 0 & 0 & \frac{\sqrt{3}}{2} & 0 \\
 \end{array}
 \right).\ee
 Its discriminant group is $G\cong\mathbb Z_3$. We choose $\lambda=(1,0,0,0,0,1/\sqrt3)^T\in \Lambda^\perp$ and $\phi$ such that $\phi(\lambda)=1$.
 This leads to the bilinear form on $G$
 \be \langle a_1|a_2\rangle={4\over 3}a_1a_2 \mod\mathbb Z\ee
 with the group weight
 \be wt(0)=0,~wt(1)=wt(2)={4\over 3}.\ee
 The evenness condition on $G\times\bar G$ is
 \be {2\over 3}(a^2-b^2)=0\mod 2\mathbb Z,\label{evene6}\ee
 therefore, any self-dual code is automatically even.
The modular $T,S$ matrices read
 \be T_{aa'}=e^{\pi i {4\over 3}a^2}\delta_{aa'} ,~~S_{aa'}={1\over \sqrt 3}e^{-2\pi i {aa'\over 3}}.\ee
 There exist $2$ even self-dual codes, generated by
 \be C_0= (11),~C_1= (12).\ee
 This leads to the modular invariants
 \be W_0=\sum_{i=0}^2 x_i\bar x_i,~W_1=\sum_{i=0}^2 x_i\bar x_{-i}.\ee
 
 Let us now calculate the Poincar\'e series of the vacuum character. The vacuum character is invariant under the congruence subgroup $\Gamma_0(3)$ and we can perform the sum as follows
 \be \hat W=\sum_{\gamma\in SL(2,\mathbb Z)/\Gamma_0(3)}\gamma(x_0\bar  x_0)=(1+\sum_{i=0}^2 T^iS)x_0\bar x_0=W_0+W_1.\ee
Therefore, the average CFT partition function is proportional to the sum over handlebodies
 \be \langle W\rangle\propto\hat W.\ee

 \subsection{$E_7$}
 
 The $E_7$ algebra has $2$ dominant weights at level $1$.
  A generator of the root lattice $\Lambda=E_7$ is
 \be \Lambda=\left(
 \begin{array}{ccccccc}
 	1 & 0 & 0 & 0 & 0 & -\frac{1}{2} & 0 \\
 	-1 & 1 & 0 & 0 & 0 & -\frac{1}{2} & 0 \\
 	0 & -1 & 1 & 0 & 0 & -\frac{1}{2} & 0 \\
 	0 & 0 & -1 & 1 & 0 & -\frac{1}{2} & 0 \\
 	0 & 0 & 0 & -1 & 1 & -\frac{1}{2} & 1 \\
 	0 & 0 & 0 & 0 & 1 & -\frac{1}{2} & -1 \\
 	0 & 0 & 0 & 0 & 0 & \frac{1}{\sqrt{2}} & 0 \\
 \end{array}
 \right),\ee
 with discriminant group $G\cong\mathbb Z_2$.
 We choose $\lambda=(1,0,0,0,0,0,1/\sqrt{2})^T\in\Lambda^\perp$ and the map $\phi$ such that $\phi(\lambda)=1$.
 This leads to the bilinear form in $G$
 \be \langle a_1|a_2\rangle={1\over 2}a_1a_2\mod \mathbb Z,\ee
 and weights
 \be wt(0)=0,~wt(1)={3\over 2} .\ee
 The evenness condition in $G\times \bar G$ is
 \be {1\over 2}(a^2-b^2)=0\mod 2\mathbb Z.\ee
The $T,S$ matrices read
 \be T_{aa'}=e^{\pi i {3\over 2}a^2}\delta_{aa'} ,\ee
 \be S_{aa'}={1\over \sqrt 2}e^{-2\pi i {1\over 2}{aa'}} .\ee
 There exists $1$ even self-dual code, generated by
 \be C= (11),\ee
 leading to the unique modular invariant CFT partition function
 \be W=x_0\bar x_0+x_1\bar x_1.\ee

\subsection{$E_8$}
 The $E_8$ lattice is unimodular, hence its discriminant group is trivial. There is a single dominant weight, with character
 \be \chi_0(\tau;\xi,t)={e^{-2\pi i t}\over (\eta(\tau))^8}\sum_{\lambda\in E_8} e^{\pi i\tau \lambda^2}e^{-2\pi i\xi\cdot\lambda} \ee
 and a unique modular invariant CFT partition function
  \be W=x_0\bar x_0.\ee

 \subsection{Ensemble average at arbitrary $n$}

 \paragraph{\underline{$E_6$}} The bilinear form on the group $G^n\times\bar G^n$ is
 \be \langle (\vec a,\vec b)|(\vec a',\vec b')\rangle={1\over 3}(\vec a\cdot\vec a'-\vec b\cdot\vec b') \mod \mathbb Z.\ee
 Due to (\ref{evene6}), any self-dual code is even. Therefore, the classification of codes is the same as in section \ref{sec:an} for $N=3$. As an application of \ref{sunalln} we conclude that 
 \be \langle W\rangle\propto\hat W.\ee

 \paragraph{\underline{$E_7$}}  The bilinear form on $G^n\times\bar G^n$ is
 \be \langle (\vec a,\vec b)|(\vec a',\vec b')\rangle={1\over 2}(\vec a\cdot\vec a'+\vec b\cdot\vec b') \mod \mathbb Z,\ee
 while the evenness condition is
 \be wt(\vec a)-wt(\vec b)={1\over 2}(\vec a^2-\vec b^2)=0\mod 2\mathbb Z.\ee
 Again, we recognize the same bilinear form and eveness condition as in section \ref{sec:an} for $N=2$. Applying \ref{sunalln}
leads to
 \be \langle W\rangle=\hat W.\ee
 
 \paragraph{\underline{$E_8$}} The discriminant group in this case is trivial. For all $n$ there exists a single code; the trivial code. This leads to a single modular invariant, given by
 \be W=x_0^n\bar x_0^n.\ee

\section{Conclusion}\label{sec:conclusions}

In this work we developed a code construction for simply-laced WZW models at level $1$. We demonstrated that the classification of the modular-invariant CFTs is equivalent to the classification of even, self-dual codes with alphabet $G=\Lambda^\perp/\Lambda$, where $\Lambda$ and $\Lambda^\perp$ are the root and weight lattices of the Lie algebra, respectively.
This formalism provides an efficient framework for constructing modular invariants and can be naturally extended to codes of larger length, corresponding to semi-simple algebras. While we focused on codes of Lorentzian signature $(n,n)$, our approach can be generalized to codes of signature $(n,m)$ or $(n,0)$ for applications to heterotic or chiral CFTs respectively.

Using this framework, we examined the holographic interpretation of these CFTs at genus 1.
Each individual CFT in our construction is holographically dual to a trivial CS theory, resulting from $U(1)^{nr}\times U(1)^{nr}$ CS theory (\ref{CS}) after gauging the subgroup of its $1$-form symmetries specified by an even, self-dual code. Moreover, the ensemble average of these CFTs is dual to summing the CS path integral over topologies. Our holographic picture is, therefore, similar to \cite{Aharony:2023zit}, but the CFTs are quite different.
We found that if all elements of the alphabet group $G$ have non-square-free order, then the uniformly-averaged boundary partition function matches the bulk CS path-integral summed over handlebody topologies only. In such cases, all even, self-dual codes belong to a single orbit of the orthogonal group $\mathcal O_n=O(n,n,G)$ (or $\mathcal O_n=O(2n,2n,\mathbb Z_2)$ for $so(4l)$), which acts on the space of codes. This leads to a unique state invariant under both $SL(2,\mathbb Z)$ and $\mathcal O_n$ (this is presumably a consequence of Howe duality \cite{dymarsky2025holographicdualityhoweduality}).
For the cases $so(4l+2)$ and $su(N)$ (when $N$ is not square-free), we find additional contributions from singular topologies. The appearance of these singular topologies can be linked to the presence of non-invertible surface operators in the chiral theory.

An interesting direction for future research is extending this code-based approach to non-simply-laced algebras and higher levels. In these cases, the fusion rules are generally non-Abelian, making a straightforward description using additive codes impossible. Instead, one could look for non-Abelian structures that capture the essential features of the special symmetric Frobenius algebra objects describing these modular invariant CFTs. 
On the other hand, the affine Lie algebras $so(2r+1)_1$ admit a free-fermion description and a formulation in terms of odd lattices—similar to the fermionic construction in Ref \cite{Kawabata:2023nlt}—may be possible. 
More broadly, modular invariants for any affine Lie algebra can, in principle, be constructed from even, self-dual lattices \cite{Gannon:1992nq,Roberts:1990tv}. However, the corresponding codes have alphabets whose size grows exponentially with the rank of the algebra, and constructing modular-invariant CFTs from them involves additional, non-trivial steps. It would be interesting to examine whether this approach has practical use.

\appendix

\section*{Acknowledgements}

The author is thankful to A. Barbar, D. Chakraborty, A. Dymarsky and A. Shapere for useful discussions. He is especially grateful to A. Dymarsky for helpful comments on the draft.


\newpage
\appendix

\renewcommand{\theequation}{A.\arabic{equation}}

\section{Even, self-dual $n=1$ codes by orbifolds\label{appOrbifold}}

The diagonal combination of characters $M_{gg'}=\delta_{g,g'}$ is always a modular invariant.  The rest of the modular invariants can be obtained by orbifolding the diagonal invariant by a subgroup $H$ of the center $\mathcal Z(\mathcal G)$ \cite{DiFrancesco:1997nk}. Since $H\subseteq\mathcal Z(\mathcal G)$ preserves the highest-weight representations, the result is a partition function of the same affine characters combined in a different way. This orbifolding procedure can be fully formulated in terms of codes, which we describe in this appendix.
 Obviously, the diagonal invariant corresponds to the diagonal code $\mathcal C=\{(g,g)|g\in G\} $. Since $G\cong \mathcal Z(\mathcal G)$, we can obtain other even, self-dual codes by ``orbifolding" the diagonal code by subgroups of its alphabet group $G$.

The isomorphism between $G$ and the center $\mathcal Z(\mathcal G)$ is straightforward; an element $g\in G$ corresponds to
\be b_g=e^{-2\pi i \lambda_g\cdot H},\ee
where $H$ are the Cartan generators in the Cartan-Weyl basis and $\phi(\lambda_g)=g$. Clearly, $b_g$ commutes with the Cartan subalgebra, while for a ladder operator $E^\alpha$ we have $b_gE^\alpha=e^{-2\pi i \lambda_g\cdot \alpha}E^\alpha b_g$. Since $\lambda_g\in \Lambda^\perp$ and $\alpha\in \Lambda$, it follows that $b_g$ commutes with all ladder operators.
The elements $b_g\in\mathcal Z(\mathcal G)$ act naturally on code polynomial variables (or chiral characters) as follows
\be b_{g'}(x_g)=e^{-2\pi i \langle g|g'\rangle}x_g.\ee
An element of the outer automorphism group $A_g\in\mathcal O_\g$ also has an action on the code variables
\be A_{g'}(x_g)=x_{g+g'}.\ee
A quick calculation shows that conjugation by the modular $S$ matrix is an explicit isomorphism between these two groups
\be A_{g'}(S(x_g))=S (b_{g'} (x_g))\implies S^\dagger A_g S=b_g.\ee

For a subgroup $H\subseteq G$, the projection onto $H$-invariant states is performed by inserting the following operator in the partition function
\be P={1\over |H|}\sum_{g\in H} b_g.\ee
The twisted sector is obtained by summing over the corresponding subgroup of the outer automorphism group $\mathcal A=\sum_{g\in H}A_g$. An extra phase arises due to the fact that $b_g$ and $A_{g'}$ do not commute, and the enumerator polynomial of the ``orbifold code" is given by \cite{DiFrancesco:1997nk}
\be W_H\equiv{1\over |H|}\sum_{g\in G} \sum_{a,a'\in H}x_g \bar x_{g+a}e^{-2\pi i \langle a'|g\rangle}e^{-\pi i \omega_a\cdot\omega_{a'}} .\ee
Orbifolding by the trivial subgroup clearly leads to the diagonal invariant, while $H=G$ results in the conjugation-symmetric invariant.
If $H$ is a proper subgroup of $G$, then $H$ is cyclic, generated by $a$, and we can write a simpler expression
\be W_a={1\over |H|}\sum_{g\in G} \sum_{p,q=0}^{|H|-1}x_g \bar x_{g+pa}e^{-2\pi i q \langle a|g\rangle}e^{-\pi i pq ~wt(a)} .\ee
Not all subgroups $H$ lead to an even, self-dual code. Rather, a subgroup generated by $a$ must satisfy 
\be {|H|\over 2}wt(a)\in\mathbb Z.\ee
\bibliographystyle{unsrt} 
\bibliography{ade} 

\end{document}